\pdfoutput=1 
\documentclass[journal]{IEEEtran}
\usepackage{breqn}
\usepackage{amsthm}
\usepackage{flushend}
\usepackage{float}
\usepackage{amsmath}
\usepackage{caption}
\usepackage{amssymb}
\usepackage{subfigure}
\usepackage{booktabs}
\usepackage{siunitx}
\usepackage{cite}
\usepackage{amsfonts,amsthm}
\usepackage{graphics, framed}
\usepackage{graphicx}
\usepackage{epsfig}
\usepackage{epstopdf}
\usepackage{mathtools}
\usepackage[linesnumbered,ruled,lined]{algorithm2e}
\usepackage{url}
\usepackage{paralist}
\usepackage[printonlyused]{acronym}
\usepackage{units}
\DeclareMathOperator*{\argmin}{arg\,min}
\usepackage{multirow}
\usepackage{multicol}
\usepackage{array, tabularx,  ragged2e,  booktabs}
\usepackage{tabu}
\usepackage{booktabs}
\usepackage[open,openlevel=1]{bookmark}
\usepackage[usenames,dvipsnames]{xcolor}
\usepackage{hyperref}
\usepackage{cleveref}

\usepackage{thmtools, thm-restate}
\usepackage{xspace}
\usepackage{color,soul}
\usepackage{balance}
\usepackage{caption}
\DeclareMathOperator*{\argmax}{argmax}
\usepackage{xcolor}
\usepackage{multirow}
\usepackage{multicol}
\usepackage{tabu}
\usepackage{booktabs}
\usepackage[open,openlevel=1]{bookmark}
\usepackage[usenames,dvipsnames]{xcolor}

\SetCommentSty{mycommfont}

\acrodef{5G}[5G]{5\textsuperscript{th}-Generation}
\acrodef{BW}[BW]{bandwidth}
\acrodef{BER}[BER]{bit error rate}
\acrodef{BPSK}[BPSK]{binary phase-shift keying}
\acrodef{CSI}[CSI]{channel state information}
\acrodef{NOMA}[NOMA]{Non-orthogonal multiple access}
\acrodef{EH}[EH]{energy harvesting}
\acrodef{BS}[BS]{base station}

\begin{document}

\title{An Optimal Joint Antenna and User Selection Algorithm for QoS-based CR-NOMA}



\author{Ömer~Faruk~Akyol, Fethi~Okta, Semiha~Tedik~Başaran~\IEEEmembership{Member,~IEEE}
  


\thanks{Ö. F. Akyol, F. Okta and S. Tedik Başaran are with the Department of Electronics, and Communications Engineering, Istanbul Technical University, 34467 Istanbul, Turkey, e-mails: \{akyolo15, okta15, tedik\}@itu.edu.tr and Ö. F. Akyol is also with HİSAR Laboratory, Informatics and Information Security
Research Center (BİLGEM), TÜBİTAK, 41470 Kocaeli, Turkey.}}

\maketitle

\begin{abstract}
Both non-orthogonal multiple access (NOMA), which can serve multiple users simultaneously and on the same frequency, and cognitive radio (CR) can contribute to eliminating the spectrum scarcity problem. In this work, an uplink CR-based NOMA (CR-NOMA) system, which is equipped with multiple users and a base station with a multi-antenna, is proposed to improve spectral efficiency. By considering the users’ quality of service (QoS), the system performance of successive interference cancellation (SIC) is investigated in this system. Two different antenna and secondary user selection algorithms are proposed to improve the outage performance and retard the effect of the error floor. The multi-antenna CR-NOMA with QoS-based SIC system outperforms conventional channel state information (CSI)-based SIC. In addition, it is shown that the outage performance of this system on both proposed algorithms is better than in the case of not using the algorithm. The closed-form outage probability expression of this system for the suboptimal joint antenna and user selection algorithm is derived. Furthermore, when the proposed algorithms are not used, the closed-form expression of the outage probability for this system with a single-antenna base station is derived. Extensive simulation results verify the accuracy of theoretical analyses.


\end{abstract}



%
\IEEEpeerreviewmaketitle

\section{Introduction}

\IEEEPARstart{I}{}n wireless communication, limited and scarce radio spectrum is an important problem, especially for the sub-6GHz spectrum. In conventional orthogonal multiple access (OMA), increasing the number of users limits orthogonal resources and the spectrum. On the other hand, non-orthogonal multiple access (NOMA) can serve multiple users efficiently by sharing the same spectrum instead of limiting the spectrum by assigning each of these users to different bands of the spectrum\cite{9802823}. Due to its ability to simultaneously serve several users while achieving great spectrum efficiency, and system capacity, and providing low complexity and a fairness/throughput tradeoff, NOMA is seen as a promising technology for the future generation of mobile communication networks \cite{elhalawany2021spectrum}. In addition, the cognitive radio (CR) concept is another promising technique to solve this radio spectrum scarcity problem. In CR, there are two different types of users, which are namely the primary user and the secondary user \cite{gao2018cognitive}. The primary user is privileged as the secondary user. Hence, the secondary user can use the spectrum only when the primary user is in the idle state \cite{hu2018full}. 

To increase the efficiency of spectrum usage, NOMA can be combined with CR~\cite{7273963}. In this way, CR-NOMA has an important potential to provide the growing requirements of users effectively in wireless communication \cite{do2020performance, li2022physical}. The authors in \cite{lv2016application} examined the outage probability of an overlay CR-based NOMA network. Also, the downlink NOMA networks combined CR over the underlay paradigm has been investigated in \cite{arzykulov2018outage}. The authors in \cite{lv2018noma} focused on a downlink CR-based NOMA system including multiple secondary users and multiple relays to enhance the outage performance. In addition, by utilizing multiple secondary users as the relay in a downlink CR-based NOMA network, the study in \cite{lv2017design} presents cooperation to CR-based NOMA. In \cite{liu2022new}, an uplink CR-based NOMA system including a rate splitting process, which provides to allocate the transmit power efficiently by the secondary user for the highest achievable rate, has been proposed. In \cite{xiang2019physical}, an overlay CR-NOMA network was analysed in terms of the physical layer security and the secrecy performance of primary users was aimed to improve. 

On the other hand, in NOMA networks, serving multiple users at the same time and frequency causes multiple-access interference. To eliminate this interference successfully, successive interference cancellation (SIC) is utilized \cite{hoang2022outage}. The decoding order of the SIC is an important issue for the performance and applicability of the system. In the literature, there are two important SIC decoding orders based on the quality of service (QoS) and the channel state information (CSI) \cite{caceres2022theoretical},\cite{ding2021new}. According to the QoS criterion, the signal of the primary user, who has higher priority QoS requirements, is decoded first. Subsequent stages of the SIC decode the signals of the other users. On the other hand, based on the CSI criteria, the signal of the users who have a better CSI than the others is first decoded, and then the remaining users are decoded using SIC following the CSI state \cite{9755045}. The outage performance results of selecting the SIC decoding orders based on both CSI and QoS in uplink NOMA systems with multi-user are investigated  \cite{ding2020unveiling},\cite{ding2020unveiling1} and the outage performances of both SIC methods in downlink NOMA systems are presented in \cite{ding2015impact}. 

However, there is no study focusing on the subject of SIC decoding orders based on either QoS or CSI for the CR-NOMA networks with multiple antennas at the receiver side. In this paper, we consider an uplink CR-NOMA system with a multi-antenna base station and multi-user. The effect of SIC decoding order on outage performance will be investigated in this proposed CR-NOMA system by comparing QoS-based SIC with CSI-based SIC. Moreover, we will select the optimal secondary user and receiver antenna to improve the outage performance of the system model through the algorithms that will be proposed in the following parts of this paper. Hence, we achieve a surprising improvement in the outage performance of the proposed NOMA system with QoS-based SIC compared with CSI-based SIC.
The main contributions of this work are given as follows:
\begin{itemize}
\item An uplink CR-NOMA network with multiple secondary users and multi-antenna is proposed. To improve the outage performance of the proposed system with QoS-based SIC, we propose two different joint antenna and user selection algorithms.
\item In the proposed system, through these proposed algorithms, the QoS-based SIC is compared to the CSI-based SIC in terms of outage probability. Numerical results show that the outage performance of the QoS-based SIC outperforms the CSI-based SIC by using these proposed algorithms. Furthermore, both proposed algorithm provides a significant gain in terms of the outage performance of this system by increasing the number of antenna or secondary user.
\item The outage probability of QoS-based SIC over the proposed algorithms is compared in both perfect CSI and imperfect CSI.
\item The closed-form expression of the outage probability of the QoS-based NOMA for the suboptimal joint antenna and user selection algorithm in the proposed system is derived and then verified with simulation results. Moreover, to show whether the optimal joint antenna and user selection algorithm works optimally in this system with QoS-based SIC, the obtained simulation results are verified by exhaustive analysis.
\end{itemize}
The rest of the paper is organized as follows. In Section II, the CR-NOMA system model is presented and this system is briefly investigated over the CSI-based SIC. In Section III, the same system model is investigated over the QoS-based SIC in detail and two algorithms are proposed to increase the outage performance of the QoS-Based SIC in this system. Section IV discusses numerical and simulation results. Finally, the conclusion is given in Section V.
\section{Conventional NOMA Network With CSI-based SIC}
\begin{figure}[!t]
    \centering
    \includegraphics[width=\linewidth]{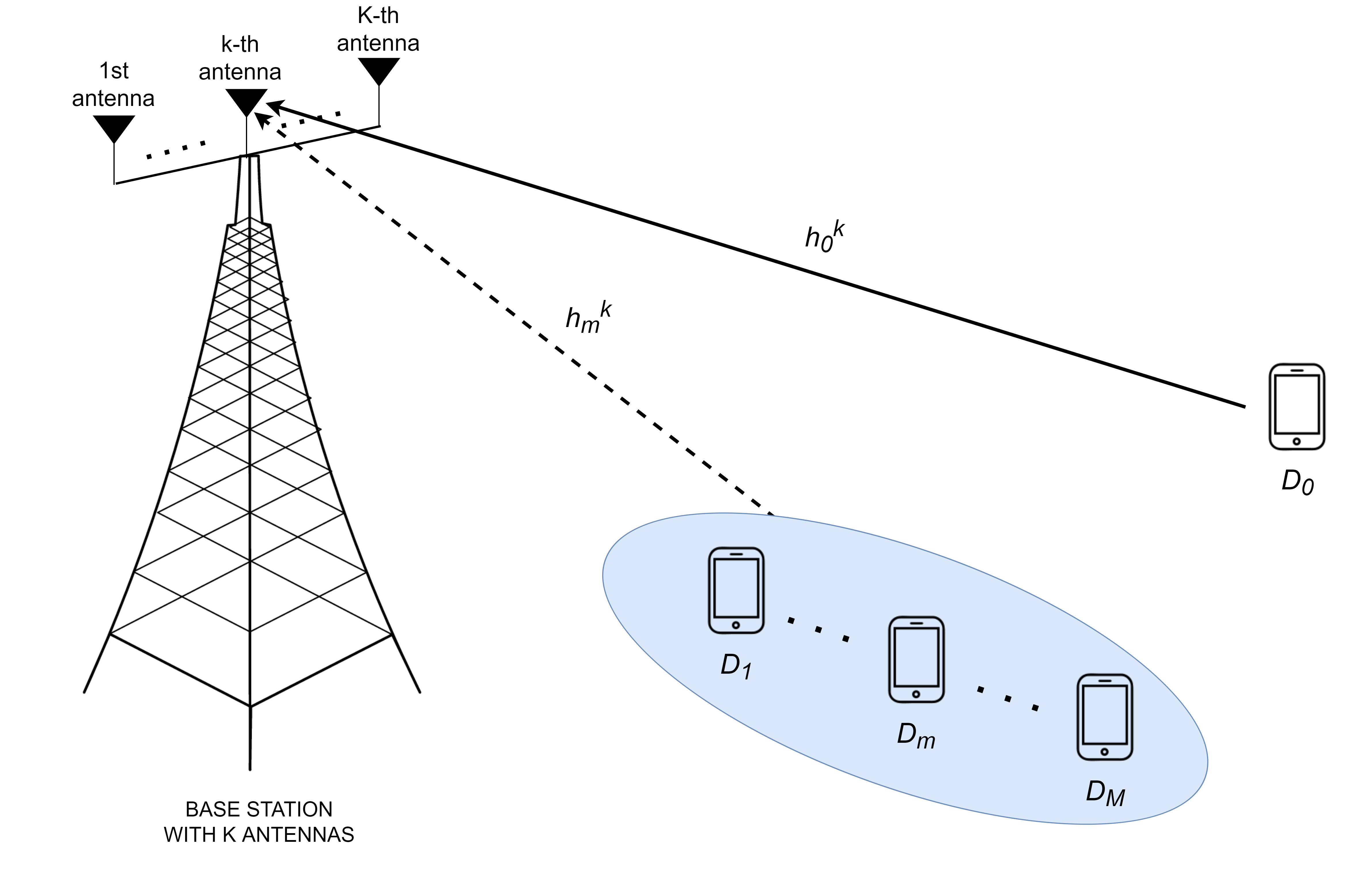}
    \caption{The considered uplink CR-NOMA system model with multi-antenna.}
    \label{fig:1}
\end{figure} 
In this section, we present the proposed multi-user multi-antenna CR-NOMA system model as shown in Fig.\ref{fig:1}. In this section, the CSI-based SIC in the proposed CR-NOMA system is investigated. In the next section, the QoS-based SIC is investigated for the same NOMA network. In this system model, there are $M+1$ users and one of these users is the primary user, stated as $D_{0}$, and the others are secondary users, stated as $D_{m}$ where $m=1,\ldots, M$. The target data rates of all secondary users and the primary user are $R_{s}^{th}$ and $R_{0}^{th}$, respectively, to detect their signals. The entire mobile user is equipped with a single antenna and the base station is equipped with $K$ antenna and $k$ is the antenna index where $k=1,\ldots, K$. The channel gains of the primary user and $m$th secondary user at the $k$th antenna of the base station are denoted by $h_{0}^k$ and $h_{m}^k$, respectively. In addition, all channels are modelled as Rayleigh fading. The mean values of the primary user's channel and the secondary users' channels are denoted by $\mathbb{E}\left[\left|h_{0}^{k}\right|^{2}\right]=\Omega_{0}$ and  $\mathbb{E}\left[\left|h_{m}^{k}\right|^{2}\right]=\Omega_{m}$, respectively, where $\mathbb{E}\left[.\right]$ is the expectation operator. In the proposed uplink CR-NOMA network, the received signal received at $k$th antenna of the base station by transmitted the information signals over $D_{0}$ and $D_{m}$ is given by
\begin{align}
    y_{S}=&\sqrt{P_{S}}h_{0}^{k}x_{PU}+\sqrt{P_{S}}h_{m}^{k}x_{SU}+w_{S},
\end{align}
where $x_{PU}$ and $x_{SU}$ are the information signals of $D_{0}$ and $D_{m}$, respectively, which these signals have a unit power in expectation. The users’ transmit powers are assumed to be identical and equal to $P_{S}$ and $w_S \sim C N\left(0, \sigma^2\right)$ is the additive
white Gaussian noise at the base station with mean zero and variance $\sigma^2$. The SIC decoding order can be determined simply by using the users’ channel conditions in the power-domain NOMA and CR-NOMA networks as in CSI-based SIC \cite{ding2020unveiling},\cite{ding2014performance}. Similar to a power-domain NOMA network serving two users with different channel conditions simultaneously, if $D_{m}$ is considered strong and $D_{0}$ weak in terms of channel conditions, firstly $D_{m}$ can be decoded at $k$th antenna of the base station by CSI-based SIC under the effect of interference by $D_{0}$ based on the following metric:
\begin{equation}
    \label{CSI_metric}
    \log \left(1+\frac{\bar{\gamma}\left|h_{m}^{k}\right|^{2}}{1+\bar{\gamma}\left|h_{0}^{k}\right|^{2}}\right) \geq R_{s}^{th}.
\end{equation}
After that, $D_{0}$ is decoded if $D_{m}$ is decoded successfully after process in (\ref{CSI_metric}). By applying CSI-based SIC process on $k$th antenna, $D_{0}$ and $D_{m}$ 's achievable data rates are given by, respectively,
\begin{align}
    R_{0}^{CSI}= &\log \left(1+\bar{\gamma}\left|h_{0}^{k}\right|^{2}\right), \\
    \label{CSI_data_rate}
    R_{m}^{CSI}= &\log \left(1+\frac{\bar{\gamma}\left|h_{m}^{k}\right|^{2}}{1+\bar{\gamma}\left|h_{0}^{k}\right|^{2}}\right), 
\end{align}
where $\bar{\gamma}=\frac{P_{S}}{\sigma^2}$ is the average signal-to-noise ratio (SNR). 
\section{NOMA Network With QoS-based SIC}
In CR-NOMA applications, selecting SIC decoding orders is frequently preferred based on the users' QoS requirements \cite{zhou2018state}. In the proposed system based on CR-NOMA, the QoS requirements of the primary user with poor channel conditions compared with the secondary users should be guaranteed \cite{7542118}. However, this system sacrifices the performance of the secondary users with better channel conditions to guarantee the primary user’s QoS requirements. Hence, the primary user's target data rate should be lower than secondary users i.e. $R_{0}^{th}<R_{s}^{th}$. Similar to an uplink NOMA network serving two users simultaneously, one of which is privileged over the other, the primary user and one of the secondary users simultaneously transmit their messages to the base station in this system model. In the $k$th antenna of the base station, as the first stage of QoS-based SIC, firstly $D_{0}$ should be decoded due to the QoS requirements under the effect of interference by $D_{m}$ based on the following metric:
\begin{equation}
    \label{QoS_metric}
    \log \left(1+\frac{\bar{\gamma}\left|h_{0}^{k}\right|^{2}}{1+\bar{\gamma}\left|h_{m}^{k}\right|^{2}}\right) \geq R_{0}^{th}.
\end{equation}
If $D_{0}$ is decoded and removed successfully after process in (\ref{QoS_metric}), $D_{m}$ can be decoded. During the QoS-based SIC process, $D_{0}$ and $D_{m}$'s achievable data rates are given by, respectively,
\begin{align}
    \label{R_0}
    R_{0}^{QoS} = & \log \left(1+\frac{\bar{\gamma}\left|h_{0}^{k}\right|^{2}}{1+\bar{\gamma}\left|h_{m}^{k}\right|^{2}}\right), \\
    R_{m}^{QoS} = &\log \left(1+\bar{\gamma}\left|h_{m}^{k}\right|^{2}\right).
\end{align}
\subsection{Joint Antenna and User Selection Algorithms}
In the proposed CR-NOMA system, to improve the outage performance of both QoS-based SIC and the entire network, user and antenna selection procedures will be suggested and applied. In the proposed CR-NOMA network with QoS-based SIC, during the first step of SIC, the primary user's signal is decoded. After the SIC process, the secondary user's signal is decoded. The outage probability of this proposed network with QoS-based SIC can be given by
\begin{equation}
\label{QoS}
    \mathrm{P}^{QoS}=\mathrm{Pr}\left(R_{0}^{QoS}<R_{0}^{th}\right)+\mathrm{Pr}\left(R_{0}^{QoS}>R_{0}^{th}, R_{m}^{QoS}<R_{s}^{th}\right),
\end{equation}
where $\mathrm{Pr}(.)$ is the probability operator. Decoding of the secondary user's signal should be guaranteed after the SIC process to improve the entire system's outage performance. Hence, both to increase the outage performance of QoS-based SIC and not restrict decoding the selected secondary user's signal at the two-stage of SIC, two algorithms are suggested, namely Algorithm~(\ref{algo_disjdecomp1}) and Algorithm~(\ref{algo_disjdecomp2}) respectively. In addition, with these proposed Algorithms, it is aimed to improve both the SIC decoding and the outage performance of the proposed system model by applying both antenna selection among $K$ antennas and user selection among $M$ secondary users. The best antenna and secondary user, which will improve both the performance of the SIC process and the performance of the entire CR-NOMA network, are selected between multiple antennas and multiple secondary users based on both algorithms.
\subsubsection{The Suboptimal Joint Antenna and User Selection Algorithm}
In Algorithm~(\ref{algo_disjdecomp1}), firstly, before starting the SIC process, the strongest data rate between the antennas and the secondary users should be compared with the target data rate of the secondary users, namely $R_{s}^{th}$, to control whether this network is completely in outage as in Line~\ref{alg:startRs}--\ref{alg:endRs}. The data rate, namely $R_{s}^{\max}$, can be given by 
\begin{align}
    \label{R_0}
    R_{s}^{\max}= \max(\max(R_{m}^{k})),
\end{align}
where $R_{m}^{k}=\log_{2}(1+\bar{\gamma}|h_{m}^{k}|^2)$. If $R_{s}^{\max}$ is bigger than $R_{s}^{th}$, the best antenna and user selection steps can be applied. However, if not, the outage occurs. In case of no outage, the best antenna is selected based on the maximum channel gain between the primary user and all antennas to improve the performance of SIC as in Line~\ref{alg:selectk}. Also, to minimize the performance degradation due to interference occurring by the secondary user in the process of SIC, which is being applied to decode the primary user's signal, the secondary user with the weakest data rate, which is greater than $R_{s}^{th}$, between the selected antenna and the secondary users is selected as in Line~\ref{alg:select_start_m}--\ref{alg:select_end_m}. Therefore, after the SIC operation, an outage in the proposed network due to the data rate of the selected secondary user being lower than $R_{s}^{th}$ is avoided. The indexes of the selected antenna and secondary user are given by 
\begin{align}
    \label{antenna_selection}
    k^{*}=&\underset{k=1,\ldots,K}{\argmax}(|h_{0}^{k}|^2), \\
    m^{*}=&\underset{R_{m}^{k^{*}}\geq R_{s}^{th}}\argmin(|h_{m}^{k^{*}}|^2).
\end{align}
Hence, the primary user's data rate is calculated according to the $k^{*}$th antenna and $m^{*}$th secondary user for the process of SIC in Line~\ref{alg:SIC1}. Before the SIC process, the primary user's data rate is given by 
\begin{align}
\label{priamry_user_data_rate}
    R_{0}^{QoS,1}=\log_{2}\Big(1+\frac{\bar{\gamma}|h_{0}^{k^{*}}|^2}{1+\bar{\gamma}|h_{m^{*}}^{k^{*}}|^2}\Big).
\end{align}
If $R_{0}^{QoS,1}$ is bigger than $R_{0}^{th}$ under the interference by the secondary user's signal, the process of SIC can be carried out successfully and the outage does not occur. If not, the outage occurs as in Line~\ref{alg:SIC_error_start}--\ref{alg:SIC_error_end}. On the other hand, if the secondary user cannot be selected since no secondary user's data rate is greater than $R_{s}^{th}$ between the $k^{*}$th antenna and all secondary users, i.e. $\max(R_{m}^{k^{*}})< R_{s}^{th}$, provided the data rate of the secondary user to be selected is greater than $R_{s}^{th}$, the channel with weakest data rate between the all secondary users and all antennas are selected. Hence, over this channel, the antenna and the secondary user are selected as in Line~\ref{alg:new_slect_m_start}--\ref{alg:new_slect_m_end}.
After that, in Line~\ref{alg:new_SIC}, the primary user's data rate is calculated according to the reselected antenna and secondary user for the process of SIC. Hence, the primary user's data rate can be expressed as
\begin{align}
    R_{0}^{QoS,2}=\log_{2}\Big(1+\frac{\bar{\gamma}|h_{0}^{k^{+}}|^2}{1+\bar{\gamma}|h_{m^{+}}^{k^{+}}|^2}\Big).
\end{align}
If $R_{0}^{QoS,2}$ is bigger than $R_{0}^{th}$ under the interference by the secondary user's signal, the process of SIC can be carried out successfully and the outage does not occur. If not, the outage occurs as in Line~\ref{alg:SIC_error_start}--\ref{alg:SIC_error_end}. Moreover, the closed-form expression of the outage probability for Algorithm~(\ref{algo_disjdecomp1}) is obtained as in (\ref{algoritms_theorical_result}) through the following procedures.
Firstly, to improve the outage performance of the proposed QoS-based system, the secondary user's signal with the weakest data rate bigger than $R_{s}^{th}$ among $M$ secondary users is selected if there is any the secondary user with the data rate bigger than $R_{s}^{th}$ in this proposed network. In case there is no secondary user with a data rate bigger than $R_{s}^{th}$, the outage occurs. Hence, the outage probability of this case can be expressed as 
\begin{equation}
    J_{1}=\mathrm{Pr}\left(R_{s}^{\max}<R_{s}^{th}\right)=F_{Y}(\gamma_{s})^{MK}=\left(1-e^{-\frac{\gamma_{s}}{\tilde{\Omega}_{m}}}\right)^{MK},
\end{equation}
where $Y=\bar{\gamma}|h_{m}^{k}|^2$. The probability density function (PDF) and the cumulative distribution function (CDF) of $Y$ are equal to $f_{Y}(y)=\frac{1}{\tilde{\Omega}_{m}}e^{-\frac{y}{\tilde{\Omega}_{m}}}$ and $F_{Y}(y)=1-e^{-\frac{y}{\tilde{\Omega}_{m}}}$, respectively. It is assumed that $\tilde{\Omega}_{0}=\bar{\gamma}\Omega_{0}$ and $\tilde{\Omega}_{m}=\bar{\gamma}\Omega_{m}$. The threshold SNRs of the primary user and each of the secondary users are $\gamma_{s}=2^{R_{s}^{th}}-1$ and $\gamma_{0}=2^{R_{0}^{th}}-1$, respectively.
After selecting the antenna with the strongest channel gain between the primary user and all antennas as in (\ref{antenna_selection}), the probability of not having any secondary users with a data rate bigger than $R_{s}^{th}$ at the selected $k^{*}$th antenna can be written by
\begin{align}
    J_{2}=&\mathrm{Pr}\left(\max\left(\log_{2}(1+\bar{\gamma}|h_{m}^{k^{*}}|^2)\right)<R_{s}^{th}\right)=F_{Y}(\gamma_{s})^{M}.
\end{align}
If there is any secondary user with a data rate bigger than $R_{s}^{th}$ at the selected antenna, in case there are $g$ secondary users with a data rate lower than $R_{s}^{th}$, the probability can be expressed using the probability mass function (PMF) as 
\begin{align}
    \mathrm{Pr}\left(G=g\right)=&\binom{M}{g}F_{Y}(\gamma_{s})^{g}\left(1-F_{Y}(\gamma_{s})\right)^{M-g}\\
    \nonumber
    =&\binom{M}{g}\left(1-e^{-\frac{\gamma_{s}}{\tilde{\Omega}_{m}}}\right)^{g}\left(e^{-\frac{\gamma_{s}}{\tilde{\Omega}_{m}}}\right)^{M-g}.
\end{align}
It is assumed that there are $M-g$ secondary users with a data rate bigger than $R_{s}^{th}$ among the selected $k^{*}$th antenna. $M-g$ out of $M$ secondary users will be distributed over the interval $\left[\gamma_{s}, \infty\right)$ and the expression of the PDF for each of these $M-g$ secondary users is $f_{\tilde{Y_{1}}}(y)=\frac{1}{\tilde{\Omega}_{m}}e^{-\frac{\left(y-\gamma_{s}\right)}{\tilde{\Omega}_{m}}}$, where $y \in\left[\gamma_{s}, \infty\right)$. Hence, to obtain the expression of the PDF for the secondary user with the minimum data rate bigger than $R_{s}^{th}$ among these $M-g$ secondary users, firstly  $\mathrm{Pr}(\tilde{Y_{1}}>y \mid G=g)$ is calculated as follows
\begin{align}
\mathrm{Pr}(\tilde{Y_{1}}>y \mid G=g)=&\left(e^{-\frac{\left(y-\gamma_{s}\right)}{\tilde{\Omega}_{m}}}\right)^{M-g}.
\end{align}
Then, $\mathrm{Pr}(\tilde{Y_{1}}>y)$ is calculated as
\begin{align}
    \mathrm{Pr}(\tilde{Y_{1}}>y)=&\mathrm{Pr}(\tilde{Y_{1}}>y \mid G=g) \mathrm{Pr}\left(G=g\right)\\
    \nonumber
    =&\left(1-e^{-\frac{\gamma_{s}}{\tilde{\Omega}_{m}}}+e^{-\frac{y}{\tilde{\Omega}_{m}}}\right)^{M}-\left(1-e^{-\frac{\gamma_{s}}{\tilde{\Omega}_{m}}}\right)^{M}.
\end{align}
Hence, the expression of $F_{\tilde{Y_{1}}}(y)$ can be written by
\begin{equation}
    F_{\tilde{Y_{1}}}(y)=1-\left(1-e^{-\frac{\gamma_{s}}{\tilde{\Omega}_{m}}}+e^{-\frac{y}{\tilde{\Omega}_{m}}}\right)^{M}+\left(1-e^{-\frac{\gamma_{s}}{\tilde{\Omega}_{m}}}\right)^{M}.
\end{equation}
Finally, the expression of $f_{\tilde{Y_{1}}}(y)$ is obtained by 
\begin{equation}
    f_{\tilde{Y_{1}}}(y)=\frac{M}{\tilde{\Omega}_{m}}e^{-\frac{y}{\tilde{\Omega}_{m}}}\left(1-e^{-\frac{\gamma_{s}}{\tilde{\Omega}_{m}}}+e^{-\frac{y}{\tilde{\Omega}_{m}}}\right)^{M-1}.
\end{equation}
Over the selected $k^{*}$th antenna and $m^{*}$th secondary user, if $R_{0}^{QoS,1}$ is lower than $R_{0}^{th}$, the outage occurs. The closed-form expression of this outage probability is calculated using $f_{\tilde{Y_{1}}}(y)$ as
\begin{align}
    \nonumber
    J_{3}=&\mathrm{Pr}\left(\frac{\bar{\gamma}|h_{0}^{k^{*}}|^2}{1+\bar{\gamma}|h_{m^{*}}^{k^{*}}|^2}<\gamma_{0}\right)
    =\int_{\gamma_{s}}^{\infty} F_{X}(\gamma_{0}+y\gamma_{0})^{K} f_{\tilde{Y_{1}}}\left(y\right) \mathrm{d} y\\
    \nonumber
    =&\frac{M}{\tilde{\Omega}_{m}}\sum_{a=0}^{K}\sum_{b=0}^{M-1}\left(-1\right)^{a} \binom{K}{a} \binom{M-1}{b} \left(1-e^{-\frac{\gamma_{s}}{\tilde{\Omega}_{m}}}\right)^{M-1-b}\\
    & \qquad \qquad \times e^{-\gamma_{s}\left(\frac{\gamma_{0}}{\tilde{\Omega}_{0}}+\frac{b+1}{\tilde{\Omega}_{m}}\right)}\frac{e^{-\gamma_{0}\frac{a}{\tilde{\Omega}_{0}}}}{\left(\frac{\gamma_{0}}{\tilde{\Omega}_{0}}+\frac{b+1}{\tilde{\Omega}_{m}}\right)},
\end{align}
where $X=\bar{\gamma}|h_{0}^{k}|^2$. The PDF and CDF of $X$ are equal to $f_{X}(x)=\frac{1}{\tilde{\Omega}_{0}}e^{-\frac{x}{\tilde{\Omega}_{0}}}$ and $F_{X}(x)=1-e^{-\frac{x}{\tilde{\Omega}_{0}}}$, respectively. If there is no secondary user with a data rate bigger than $R_{s}^{th}$ at the selected $k^{*}$th antenna, provided there is any secondary user with a data rate bigger than $R_{s}^{th}$ for any antenna, the channels, which can provide a data rate bigger than $R_{s}^{th}$ between the secondary users and the antennas, are checked and detected. Among these detected channels, the one with the lowest data rate is selected. Hence, both the antenna and the secondary user are selected over the selected channel. The probability of having $v$ channels, which provide a data rate lower than $R_{s}^{th}$ between the secondary users and the antennas, can be expressed as
\begin{align}
    \mathrm{Pr}\left(V=v\right)
    =&\binom{KM}{v}\left(1-e^{-\frac{\gamma_{s}}{\tilde{\Omega}_{m}}}\right)^{v}\left(e^{-\frac{\gamma_{s}}{\tilde{\Omega}_{m}}}\right)^{KM-v}.
\end{align}
Hence, it is assumed that there are $KM-v$ channels with a data rate bigger than $R_{s}^{th}$ between the secondary users and the antennas. $KM-v$ out of $KM$ channels will be distributed over the interval $\left[\gamma_{s}, \infty\right)$ and the expression of the PDF for each of these $KM-v$ channels is $f_{\tilde{Y_{2}}}(y)=\frac{1}{\tilde{\Omega}_{m}}e^{-\frac{\left(y-\gamma_{s}\right)}{\tilde{\Omega}_{m}}}$, where $y \in\left[\gamma_{s}, \infty\right)$. To obtain the expression of PDF for the channel that will provide the minimum data rate bigger than $R_{s}^{th}$ among $KM-v$ channels, $\mathrm{Pr}(\tilde{Y_{2}}>y \mid V=v)$ is calculated as follows
\begin{align}
\mathrm{Pr}(\tilde{Y_{2}}>y \mid V=v)=&\left(e^{-\frac{\left(y-\gamma_{s}\right)}{\tilde{\Omega}_{m}}}\right)^{KM-v}.
\end{align}
Hence, $\mathrm{Pr}(\tilde{Y_{2}}>y)$ is calculated using $\mathrm{Pr}(\tilde{Y_{2}}>y \mid V=v)$ and $\mathrm{Pr}\left(V=v\right)$ as
\begin{align}
    \mathrm{Pr}(\tilde{Y_{2}}>y)=&\mathrm{Pr}(\tilde{Y_{2}}>y \mid V=v) \mathrm{Pr}\left(V=v\right)\\
    \nonumber
    =&\left(1-e^{-\frac{\gamma_{s}}{\tilde{\Omega}_{m}}}+e^{-\frac{y}{\tilde{\Omega}_{m}}}\right)^{KM}-\left(1-e^{-\frac{\gamma_{s}}{\tilde{\Omega}_{m}}}\right)^{KM}.
\end{align}
The CDF expression of $F_{\tilde{Y_{2}}}(y)$ can be given by
\begin{equation}
    F_{\tilde{Y_{2}}}(y)=1-\left(1-e^{-\frac{\gamma_{s}}{\tilde{\Omega}_{m}}}+e^{-\frac{y}{\tilde{\Omega}_{m}}}\right)^{KM}+\left(1-e^{-\frac{\gamma_{s}}{\tilde{\Omega}_{m}}}\right)^{KM}.
\end{equation}
Finally, the PDF expression of $f_{\tilde{Y_{2}}}(y)$ is obtained by 
\begin{equation}
    f_{\tilde{Y_{2}}}(y)=\frac{KM}{\tilde{\Omega}_{m}}e^{-\frac{y}{\tilde{\Omega}_{m}}}\left(1-e^{-\frac{\gamma_{s}}{\tilde{\Omega}_{m}}}+e^{-\frac{y}{\tilde{\Omega}_{m}}}\right)^{KM-1}
\end{equation}
Over the selected $k^{+}$th antenna and $m^{+}$th secondary user, if $R_{0}^{QoS,2}$ is lower than $R_{0}^{th}$, the outage occurs. The closed-form expression of this outage probability is calculated using $f_{\tilde{Y_{2}}}(y)$ as
\begin{align}
    J_{4}=&\mathrm{Pr}\left(\frac{\bar{\gamma}|h_{0}^{k^{+}}|^2}{1+\bar{\gamma}|h_{m^{+}}^{k^{+}}|^2}<\gamma_{0}\right)=\int_{\gamma_{s}}^{\infty} F_{X}(\gamma_{0}+y\gamma_{0}) f_{\tilde{Y_{2}}}\left(y\right) \mathrm{d} y \\
    \nonumber
    =&\frac{KM}{\tilde{\Omega}_{m}}\sum_{c=0}^{1}\sum_{d=0}^{KM-1}\left(-1\right)^{c} \binom{KM-1}{d} e^{-\gamma_{s}\left(\frac{\gamma_{0}}{\tilde{\Omega}_{0}}+\frac{d+1}{\tilde{\Omega}_{m}}\right)} \\
    \nonumber
    &\qquad \qquad  \times \left(1-e^{-\frac{\gamma_{s}}{\tilde{\Omega}_{m}}}\right)^{KM-1-d} \frac{e^{-\gamma_{0}\frac{c}{\tilde{\Omega}_{0}}}}{\left(\frac{\gamma_{0}}{\tilde{\Omega}_{0}}+\frac{d+1}{\tilde{\Omega}_{m}}\right)}.
    \nonumber
\end{align}
Finally, the closed-form expression of the outage probability for Algorithm~(\ref{algo_disjdecomp1}) can be written as
\begin{equation}
    \label{final_outage_closed}
    \mathrm{P}^{QoS}=J_{1}+\left(1-J_{1}\right)\left[\left(1-J_{2}\right)J_{3}+J_{2}J_{4}\right].
\end{equation}  
Using (\ref{final_outage_closed}), the closed-form theoretical expression of Algorithm~(\ref{algo_disjdecomp1}) is given in (\ref{algoritms_theorical_result}). On the other hand, in case the base station with a single antenna, the secondary user with minimum channel gain among $M$ secondary user is selected to improve the outage performance of the NOMA network with QoS-based SIC as in \cite{ding2020unveiling}. Therefore, the outage performances of Algorithm~(\ref{algo_disjdecomp1}) and secondary user selection method for QoS-based SIC in \cite{ding2020unveiling} are investigated in the numerical results section in the case of a single antenna base station.
\begin{table*}[!t]
\centering
\begin{minipage}{\textwidth}
\begin{equation}
\begin{aligned}
\label{algoritms_theorical_result}
    \mathrm{P}^{QoS}= \left(1-\left(1-e^{-\frac{\gamma_{s}}{\tilde{\Omega}_{m}}}\right)^{MK}\right) \Bigg[\left(1-\left(1-e^{-\frac{\gamma_{s}}{\tilde{\Omega}_{m}}}\right)^{M}\right) \frac{M}{\tilde{\Omega}_{m}}\sum_{a=0}^{K}\sum_{b=0}^{M-1}\left(-1\right)^{a} \binom{K}{a} \binom{M-1}{b} \left(1-e^{-\frac{\gamma_{s}}{\tilde{\Omega}_{m}}}\right)^{M-1-b} \frac{e^{-\gamma_{0}\frac{a}{\tilde{\Omega}_{0}}}}{\left(\frac{\gamma_{0}}{\tilde{\Omega}_{0}}+\frac{b+1}{\tilde{\Omega}_{m}}\right)} e^{-\gamma_{s}\left(\frac{\gamma_{0}}{\tilde{\Omega}_{0}}+\frac{b+1}{\tilde{\Omega}_{m}}\right)}\\
    +\left(1-e^{-\frac{\gamma_{s}}{\tilde{\Omega}_{m}}}\right)^{M} \frac{KM}{\tilde{\Omega}_{m}}\sum_{c=0}^{1}\sum_{d=0}^{KM-1}\left(-1\right)^{c} \binom{KM-1}{d} e^{-\gamma_{s}\left(\frac{\gamma_{0}}{\tilde{\Omega}_{0}}+\frac{d+1}{\tilde{\Omega}_{m}}\right)} \left(1-e^{-\frac{\gamma_{s}}{\tilde{\Omega}_{m}}}\right)^{KM-1-d} \frac{e^{-\gamma_{0}\frac{c}{\tilde{\Omega}_{0}}}}{\left(\frac{\gamma_{0}}{\tilde{\Omega}_{0}}+\frac{d+1}{\tilde{\Omega}_{m}}\right)}\Bigg]+ \left(1-e^{-\frac{\gamma_{s}}{\tilde{\Omega}_{m}}}\right)^{MK}
\end{aligned}
\end{equation}
\medskip
\hrule
\end{minipage}
\end{table*}
\begin{algorithm}[t]
\caption{The Suboptimal Joint Antenna and User Selection Algorithm}
\label{algo_disjdecomp1}

    \SetKwInOut{Input}{input}
    \SetKwInOut{Output}{output}

    \Input{$K$, $M$, $\mathcal{G}[M,K-1]$, $\mathcal{L}[1...K]$}
    \Output{$R^{QoS}$, $error$}

    \eIf{\label{alg:startRs}$R_{s}^{\max}<R_{s}^{th}$}
    {
    $error$\;\label{alg:endRs}
    }{
    $k^{*} \gets\underset{k=1,\ldots,K}{\mathrm{argmax}}(|h_{0}^{k}|^2)$\label{alg:selectk}\\
    \eIf{\label{alg:select_start_m}$\max(R_{m}^{k^{*}})\geq R_{s}^{th}$}{
    
        \For {$m \gets1$ \KwTo $M$}{
            $\mathcal{L}[m] \gets 0$\\
            \If{$R_{m}^{k^{*}}\geq R_{s}^{th}$} {$\mathcal{L}[m] \gets |h_{m}^{k^{*}}|^2$}
        } 
    
        $m^{*} \gets \underset{m}{\mathrm{argmin}}(\mathcal{L}|\mathcal{L}[m]>0)$\label{alg:select_end_m}
        
        $R^{QoS} = \log_{2}\Big(1+\frac{\bar{\gamma}|h_{0}^{k^{*}}|^2}{1+\bar{\gamma}|h_{m^{*}}^{k^{*}}|^2}\Big)$\label{alg:SIC1}
    }    
    {
    \label{alg:new_slect_m_start}
    \For {$m \gets1$ \KwTo $M$}{
        
        \For {$k \gets1$ \KwTo $K$ / $k^{*}$}{
            $\mathcal{G}[m,k] \gets 0$\\
        \If{$R_{m}^{k}\geq R_{s}^{th}$} {$\mathcal{G}[m,k] \gets |h_{m}^{k}|^2$}
        
        }
        
    }
    
    $m^{+},k^{+} \gets \underset{m}{\mathrm{argmin}} \underset{k} {\mathrm{argmin}}(\mathcal{G}|\mathcal{G}[m,k]>0)$\label{alg:new_slect_m_end}
    
    $R^{QoS}=\log_{2}\Big(1+\frac{\bar{\gamma}|h_{0}^{k^{+}}|^2}{1+\bar{\gamma}|h_{m^{+}}^{k^{+}}|^2}\Big)$\label{alg:new_SIC}
    
    }
    \If{$R^{QoS}<R_{0}^{th}$}{\label{alg:SIC_error_start}
        $error$\;
    \label{alg:SIC_error_end}}
    }

\end{algorithm}
\subsubsection{The Optimal Joint Antenna and User Selection Algorithm}
In another proposed algorithm for the CR-NOMA network with QoS-based SIC, firstly, as at the beginning of the Algorithm~(\ref{algo_disjdecomp1}), to control if this network is in the outage completely before the SIC process, it should be checked whether $R_{s}^{\max}$ is bigger than $R_{s}^{th}$. If $R_{s}^{\max}$ is bigger than $R_{s}^{th}$, the antenna and user selection stages can be applied. If not, it is accepted that the outage occurs in this network before the process of SIC as in Line~\ref{alg:algoritma2_Rs_control_start}--\ref{alg:algoritma2_Rs_control_end}. In the antenna and user selection stages, all channels between secondary users and antennas, satisfying the condition that the data rate is bigger than $R_{s}^{th}$, are detected in Line~\ref{alg:algoritma2_channel_start}--\ref{alg:algoritma2_channel_end}. After that, all possible antennas, which can be used to select the best antenna, are determined by controlling whether there is any secondary user with a data rate greater than $R_{s}^{th}$ between each antenna and secondary users as in Line~\ref{alg:algoritma2_secondary_user_selection1_start}. The secondary users with a data rate greater than $R_{s}^{th}$ but the lowest data rate among $M$ secondary users is detected for each of these possible antennas separately as in Line~\ref{alg:algoritma2_secondary_user_selection_m}. Using the secondary users detected separately for each of these possible antennas, as many as the number of all possible antennas that can be selected, the expression of $|h_{0}^{k}|^2/|h_{m^{\star}}^{k}|^2$ is repeatedly calculated and recorded in Line~\ref{alg:algoritma2_antenna_selection_k}. By using these recorded values, the best antenna and accordingly the best secondary user are selected as in Line~\ref{alg:algoritma2_antenna_s}. After the stage of antenna and secondary user selection, the process of SIC will be applied. The primary user's data rate is given by
\begin{align}
    R_{0}^{QoS,3}=\log_{2}\Big(1+\frac{\bar{\gamma}|h_{0}^{k^{\star}}|^2}{1+\bar{\gamma}|h_{m^{\star}}^{k^{\star}}|^2}\Big).
\end{align}
If $R_{0}^{QoS,3}$, which is calculated in Line~\ref{alg:algoritma2_SIC}, is greater than $R_{0}^{th}$, the process of SIC is carried out successfully by being decoded the primary user's signal. If not, the outage occurs in the proposed NOMA network as in Line~\ref{alg:algoritma2_outage_start}--\ref{alg:algoritma2_outage_end}. 
\begin{algorithm}[!t]
\caption{The Optimal Joint Antenna and User Selection Algorithm}
\label{algo_disjdecomp2}

    \SetKwInOut{Input}{input}
    \SetKwInOut{Output}{output}

    \Input{$K$, $M$, $\mathcal{L}[1...M]$, $\mathcal{H}[1...K]$}
    \Output{$R^{QoS}$, $error$}
    
\eIf{\label{alg:algoritma2_Rs_control_start}$R_{s}^{\max}<R_{s}^{th}$}
{
    $error$\;
\label{alg:algoritma2_Rs_control_end}}{

    {\For {$k \gets1$ \KwTo $K$}{
        $Sum \gets 0$\\
        $\mathcal{H}[k] \gets 0$\\
        \For {$m \gets1$ \KwTo $M$}{
            $\mathcal{L}[m] \gets 0$\\
            \If{$R_{m}^{k}\geq R_{s}^{th}$\label{alg:algoritma2_channel_start}}{$\mathcal{L}[m] \gets |h_{m}^{k}|^2$\label{alg:algoritma2_channel_end}\\
            $Sum \gets Sum+1$\\
            \Else{$Sum \gets Sum+0$\\}}
            }\If{\label{alg:algoritma2_secondary_user_selection1_start}$Sum > 0$}{
            $m^{\star} \gets \underset{m }{\arg\min}(\mathcal{L}|\mathcal{L}[m]>0)$\label{alg:algoritma2_secondary_user_selection_m}\\
            $\mathcal{H}[k]\gets\label{alg:algoritma2_antenna_selection_k} |h_{0}^{k}|^2/|h_{m^{\star}}^{k}|^2$   \label{alg:algoritma2_secondary_user_selection1_end}
            }
        }
    }
    $k^{\star} \gets \underset{k }{\arg\max}(\mathcal{H})$\label{alg:algoritma2_antenna_s}
    
    $R^{QoS} \gets \log_{2}\Big(1+\frac{\bar{\gamma}|h_{0}^{k^{\star}}|^2}{1+\bar{\gamma}|h_{m^{\star}}^{k^{\star}}|^2}\Big)$ \label{alg:algoritma2_SIC}
     
    \If{\label{alg:algoritma2_outage_start}$R^{QoS}<R_{0}^{th}$}{
        $error$\;
     \label{alg:algoritma2_outage_end}   
    }
}
\end{algorithm}
\subsection{QoS-based SIC with imperfect channel estimation}
In this subsection, we present the channel model that will be used to examine how well the suggested network with imperfect CSI performs in channels with independent and identically distributed (i.d.d.) Rayleigh fading. Assuming the channel estimation is imperfect, the primary pilot symbol assisted channel estimate $\hat{h}_{0}^{k}$, which is between the primary user and $k$th antenna, and the secondary pilot symbol assisted channel estimate $\hat{h}_{m}^{k}$, which is between the $m$th secondary user and $k$th antenna, differ from true channels $h_{0}^{k}$ and $h_{m}^{k}$ by an independent complex Gaussian error $\Delta h_{0}^{k}$ and $\Delta h_{m}^{k}$, respectively, which are with zero-mean and variance $\sigma_{e}^{2}$. Hence, $\hat{h}_{0}^{k}$ and $\hat{h}_{m}^{k}$ are equal to $h_{0}^{k} + \Delta h_{0}^{k}$ and $h_{m}^{k} + \Delta h_{m}^{k}$, respectively \cite{1413621}. Assuming $\Omega_{0}=\Omega_{m}=1$, the estimate channels $\hat{h}_{0}^{k}$ and $\hat{h}_{m}^{k}$ are i.i.d. complex Gaussian random variables with zero-mean and variance $1+\sigma_{e}^{2}$. The true channel gains $h_{0}^{k}$ and $h_{m}^{k}$ can be written in terms of $\hat{h}_{0}^{k}$ and $\hat{h}_{m}^{k}$, respectively \cite{gu2003performance}, 
\begin{align}
    h_{0}^{k} &= \Gamma \hat{h}_{0}^{k} + \tilde{h}_{0}^{k}\\
    \nonumber
    h_{m}^{k} &= \Gamma \hat{h}_{m}^{k} + \tilde{h}_{m}^{k}
\end{align}
where $\Gamma = \frac{1}{1+\sigma_{e}^{2}}$. Also, $\tilde{h}_{0}^{k}$ and $\tilde{h}_{m}^{k}$ are i.i.d. complex Gaussian random variables with zero-mean and variance $\tilde{\sigma}^{2}=\frac{\sigma_{e}^{2}}{1+\sigma_{e}^{2}}$.
\section{Numerical And Simulation Results}
\begin{figure}[!t]
\includegraphics[width=\linewidth]{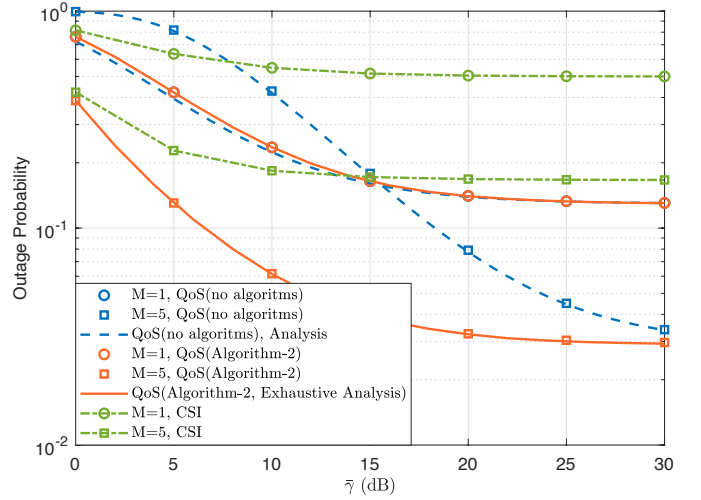}
\caption{The outage probability against SNR for different number of the secondary users. The users’ channel gains are assumed to be i.i.d. Rayleigh fading. $R_{0}^{th}$ = 0.2 bit per channel uses (BPCU), $R_{s}^{th}$ = 1 BPCU, $\Omega_{0}=\Omega_{m}=1$ and $K=1$.}
\label{fig:2}
\vskip-0.5cm
\centering
\end{figure}

\begin{figure}[!t]
\includegraphics[width=\linewidth]{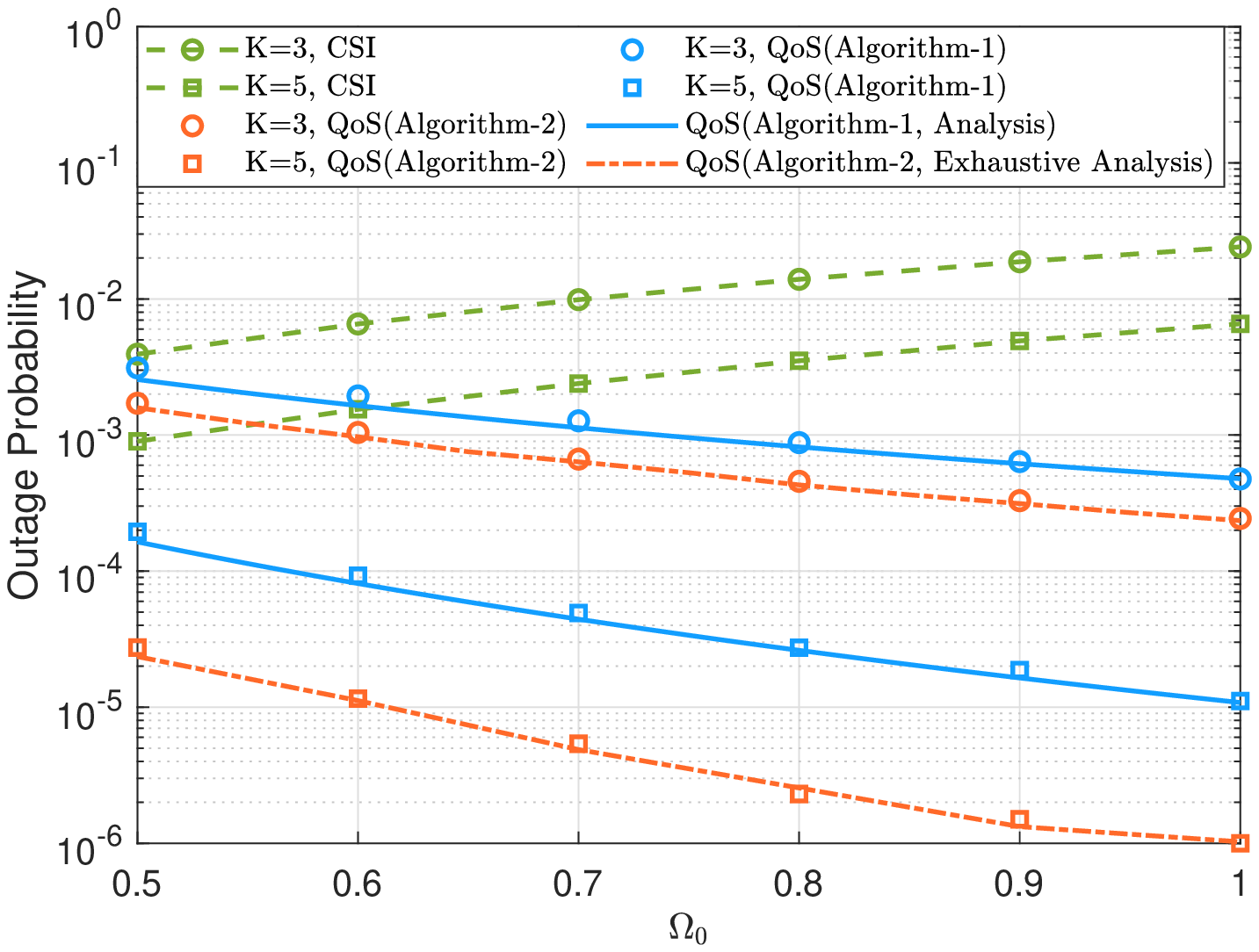}
\caption{The outage probability against $\Omega_{0}$ for different number of antennas. The users’ channel gains are assumed to be i.i.d. Rayleigh fading. $R_{0}^{th}$ = 0.2 BPCU, $R_{s}^{th}$ = 1 BPCU, $\Omega_{m}=1$, $\bar{\gamma}$ = 10dB, and $M=5$.}
\label{fig:6}
\vskip-0.5cm
\centering
\end{figure}

In this section, for the proposed uplink CR-NOMA scenario, the QoS-based SIC is compared with the CSI-based SIC in terms of outage probability over two different algorithm methods and then, in case both perfect channel state and imperfect channel state, the outage performance of the QoS-based SIC over these algorithms is investigated according to the different number of antennas and secondary users. In addition, the outage performance of the QoS-based SIC for Algorithm~(\ref{algo_disjdecomp1}) is evaluated with numerical results obtained by Monte Carlo simulations and these numerical results are verified by theoretical results. Also, the outage performance of the QoS-based SIC for Algorithm~(\ref{algo_disjdecomp2}) is confirmed by the exhaustive analysis that is a primary method to find the optimal scheduled secondary user and antenna \cite{7510916}. On the other hand, to present the performance of the CSI-based SIC for this proposed system in Fig.~\ref{fig:2} and Fig.~\ref{fig:6}, the antenna with the weakest channel gain between the primary user and antennas is selected to increase the data rate of the secondary user as in (\ref{CSI_data_rate}) and, then the secondary user with the strongest channel gain between the selected antenna and secondary users is selected.
In Fig.\ref{fig:2}, the impact of $M$ on the outage probability versus SNR is investigated for the proposed system where $[M]:[1,5]$ and $[K]:[1]$. In case $K$ is 1, as well as increasing $M$ degrades the outage performance, it is seen that the outage performance of QoS-based SIC with and without algorithms is better compared to CSI-based SIC. Moreover, the effect of the error floor in CSI-based SIC is dramatically seen compared to QoS-based SIC with and without algorithms. In case $M$ is 5, without the use of any proposed algorithms, the outage probability of CSI-based SIC is obtained by selecting the secondary user with maximum channel gain between the antenna and secondary users, while the outage probability of QoS-based SIC is obtained by selecting the secondary user with minimum channel gain between the antenna and secondary users \cite{ding2020unveiling}. 
It can be seen that the analysis results and the exhaustive analysis are in agreement with the simulation results. The simulation results of the outage probability of the QoS-based SIC without the proposed algorithms are verified through the closed-form expression of $\mathrm{P}_{A}^{QoS}$ obtained in the Appendix.
\begin{figure}[!t]
\includegraphics[width=\linewidth]{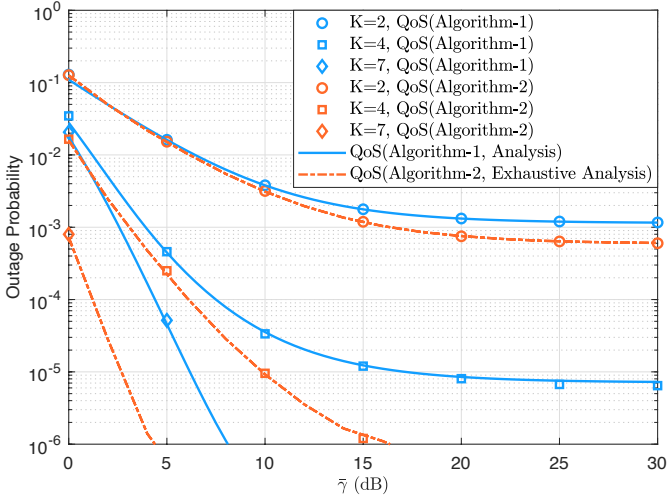}
\caption{The outage probability against SNR for different number of antennas. The users’ channel gains are assumed to be i.i.d. Rayleigh fading. $R_{0}^{th}$ = 0.2 BPCU, $R_{s}^{th}$ = 1 BPCU, $\Omega_{0}=\Omega_{m}=1$, and $M=6$.}
\label{fig:3}
\vskip-0.5cm
\centering
\end{figure}
Fig.\ref{fig:3} depicts the impact of the number of antennas in the base station on the outage probability versus SNR for the proposed system where $[M]:[6]$ and $[K]:[2,4,7]$. In case $M$ is 6, it can be seen that increasing $K$ dramatically degrades the outage probability. When $K$ is 2, although it is seen that the error floor begins to occur at high SNR, the effect of the error floor decreases as the number of antennas increases. The outage performance of QoS-based SIC with Algorithm~(\ref{algo_disjdecomp2}) is better compared with Algorithm~(\ref{algo_disjdecomp1}) especially. It is seen that the simulation results are verified by the theoretical analysis for Algorithm 1 and the exhaustive analysis for Algorithm 2.

\begin{figure}[!t]
\includegraphics[width=\linewidth]{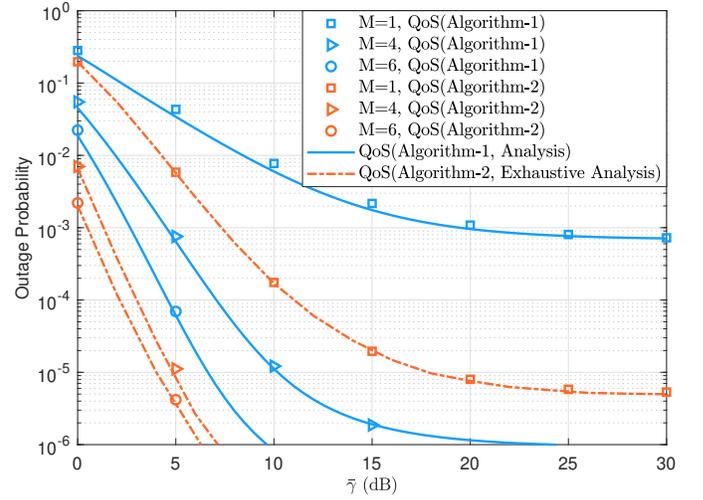}
\caption{The outage probability against SNR for different number of the secondary users. The users’ channel gains are assumed to be i.i.d. Rayleigh fading. $R_{0}^{th}$ = 0.2 BPCU, $R_{s}^{th}$ = 1 BPCU, $\Omega_{0}=\Omega_{m}=1$, and $K=6$.}
\label{fig:4}
\vskip-0.5cm
\centering
\end{figure}

In Fig.\ref{fig:4}, the impact of $M$ on the outage probability versus SNR is investigated for the proposed system where $[M]:[1,4,6]$ and $[K]:[6]$. In case $K$ is 6, as well as increasing $M$ degrades the outage performance, it is seen that the effect of the error floor is significantly reduced. Also, increasing $M$ dramatically degrades the outage probability for both Algorithms. As in Fig. \ref{fig:3}, the outage performance of QoS-based SIC with Algorithm~(\ref{algo_disjdecomp2}) is better than compared with Algorithm~(\ref{algo_disjdecomp1}). Moreover, it can be seen that the analysis results and the exhaustive analysis are in agreement with the simulation results. 

\begin{figure}[!t]
\includegraphics[width=\linewidth]{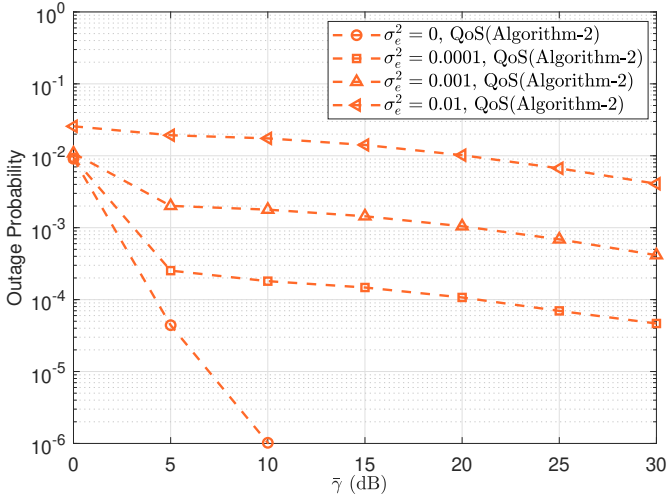}
\caption{The outage probability against SNR in case of imperfect CSI. The users’ channel gains are assumed to be i.i.d. Rayleigh fading. $R_{0}^{th}$ = 0.2 BPCU, $R_{s}^{th}$ = 1 BPCU, $\Omega_{0}=\Omega_{m}=1$, and $K=M=4$.}
\label{fig:5}
\vskip-0.5cm
\centering
\end{figure}

In Fig.\ref{fig:5}, the effect of $\sigma_{e}^2$ in case imperfect CSI in the outage probability against SNR is investigated compared to perfect CSI for the proposed system where $K$ and $M$ is 5. In the case of imperfect CSI, the outage probability in the NOMA network with the QoS-based SIC increases for Algorithm~(\ref{algo_disjdecomp2}) compared to the perfect CSI case. Although there is a channel estimation error in the case of imperfect CSI, it is observed that Algorithm~(\ref{algo_disjdecomp2}) works quite well.
In Fig.\ref{fig:6}, in case of bad channel conditions between primary user and antennas compared to secondary users, the outage probability against $\Omega_{0}$ for the proposed system is investigated, where $[M]:[5]$ and $[K]:[3, 5]$. As $\Omega_{0}$ increases, the outage probability in the NOMA network with the QoS-based SIC for both proposed algorithms increases compared to CSI-based SIC. As in Fig. \ref{fig:4}, the outage performance of QoS-based SIC with Algorithm~(\ref{algo_disjdecomp2}) is better than compared with Algorithm~(\ref{algo_disjdecomp1}). Increasing the number of antennas significantly improves the outage performance of QoS-based SIC compared to CSI-based SIC. In case $K$ is 3, the outage performance of the QoS-based SIC is approximately equal to the outage performance of CSI-based SIC for both proposed algorithms when $\Omega_{0}$ is about 0.5. However, in case $K$ is 5, the outage performance of the QoS-based SIC for both proposed algorithms is equal to the outage performance of CSI-based SIC for both proposed algorithms when $\Omega_{0}$ is less than 0.5. It can be seen that the analysis results and the exhaustive analysis are in agreement with the simulation results.

\section{Conclusion}
In this paper, we have proposed an uplink CR-NOMA system with QoS-based SIC consisting of multiple users and the base station with a multi-antenna. The outage performance of this system is aimed to improve by selecting the antenna and the secondary user over two different proposed algorithms for QoS-based SIC. Also, the effect of the error floor is aimed to decrease for this system with QoS-based SIC by selecting the antenna and the secondary user in case of multi-user multi-antenna. On the other hand, to observe the effect of selecting SIC decoding order on the system performance, the outage performance of QoS-based SIC through these algorithms has been compared with CSI-based SIC over the proposed system. The exact expression of the outage probability of this proposed system model with QoS-based SIC for Algorithms~(\ref{algo_disjdecomp1}) has been derived in closed form over Rayleigh fading channels. In addition, the outage performance of this system with QoS-based SIC for Algorithms~(\ref{algo_disjdecomp2}) has been verified by the exhaustive analysis. In the case base station with a single antenna, the closed-form expression of the outage probability of this system with QoS-based SIC by not its use of proposed algorithms has also been derived and compared with proposed algorithms. In the case of multi-antenna in the base station, the proposed algorithms for QoS-based SIC have provided to degrade the outage probability effectively. Also, the outage probability is degraded by increasing the number of secondary users and antenna and the error floor is significantly decreased. Increasing the number of antennas results in a more effective increase in the outage performance difference between Algorithms~(\ref{algo_disjdecomp1}) and Algorithms~(\ref{algo_disjdecomp2}) compared to increasing the number of secondary users. 
\vspace{-1.99mm}
\renewcommand{\theequation}{A.\arabic{equation}}
\setcounter{equation}{0}  
\section*{Appendix}
Utilizing the secondary user signal with the weakest channel gain among $M$ secondary users simply to minimize the performance degradation of the primary user if the base station has only one antenna in the proposed network, the expression of outage probability can be written as 
\begin{equation}
\label{j1_and_j2}
    \resizebox{.99\hsize}{!}{$\mathrm{P}_{A}^{QoS}=\underbrace{\mathrm{Pr}\left(\frac{X}{1+Y_{1}}<\gamma_{0}\right)}_{K_{1}}+\underbrace{\mathrm{Pr}\left(\frac{X}{1+Y_{1}}>\gamma_{0}, Y_{1}<\gamma_{s}\right)}_{K_{2}}$},
\end{equation}
where $X=\bar{\gamma}|h_{0}|^{2}$ and $Y_{m}=\bar{\gamma}|h_{m}|^{2}$. Also, the PDFs of the unordered variable $X$ and $Y$ are $f_{X}(x)=\frac{1}{\tilde{\Omega}_{0}}e^{-\frac{x}{\tilde{\Omega}_{0}}}$, $f_{Y_{m}}(y)=\frac{1}{\tilde{\Omega}_{m}}e^{-\frac{y}{\tilde{\Omega}_{m}}}$, respectively. It is assumed that $|h_{1}|^{2}\leq |h_{m}|^{2}\leq|h_{M}|^{2}$ in this part. In what follows, $K_{1}$ and $K_{2}$ are addressed to write the closed-form expression of $\mathrm{P}_{A}^{QoS}$ given in (\ref{j1_and_j2}). Firstly, the expression of $K_{1}$ can be given as
\begin{equation}
    \label{J1}
    K_{1}=\int_{0}^{\infty} F_{X}\left(\gamma_{0}+\gamma_{0}y\right) f_{Y_{1}}(y) \mathrm{d} y.
\end{equation}
To calculate the expression of $K_{1}$, $F_{X}\left(\gamma_{0}+\gamma_{0}y\right)$ and $f_{Y_{1}}(y)$ need to be evaluated first. Firstly, $F_{X}\left(\gamma_{0}+\gamma_{0}y\right)$ is equal to $1-e^{-\frac{\left(y+1\right)\gamma_{0}}{\tilde{\Omega}_{0}}}$. With the aid of \cite{men2016performance}, the PDF of the ordered variable $Y_{1}$ can be calculated from the CDF of the ordered variable $Y_{1}$, $F_{Y_{1}}(y)$ is given by 
\begin{equation}
    \label{CDF_y1}
    F_{Y_{1}}(y)=M \sum_{s=0}^{M-1} \frac{\left(-1\right)^{s}}{1+s}\binom{M-1}{s} \left(1-e^{-\frac{y}{\tilde{\Omega}_{1}}}\right)^{s+1},
\end{equation}
where $\tilde{\Omega}_{1}=\bar{\gamma}\Omega_{1}$. Using (\ref{CDF_y1}), the PDF of the ordered variable $Y_{1}$ can be written as 
\begin{align}
    f_{Y_{1}}(y)&= \frac{M}{\tilde{\Omega}_{1}}\sum_{s=0}^{M-1} \left(-1\right)^{s} \binom{M-1}{s} e^{-\frac{y}{\tilde{\Omega}_{1}}} \left(1-e^{-\frac{y}{\tilde{\Omega}_{1}}}\right)^{s}.
\end{align}
Hence, $K_{1}$ can be rewritten as
\begin{align}
\label{J1_closed}
    \nonumber
    K_{1}&=1-e^{-\frac{\gamma_{0}}{\tilde{\Omega}_{1}}}\int_{0}^{\infty} e^{-\frac{-y \gamma_{0}}{\tilde{\Omega}_{1}}} f_{Y_{1}}(y) \mathrm{d} y\\
    &=1-\frac{M}{\tilde{\Omega}_{1}} e^{-\frac{\gamma_{0}}{\tilde{\Omega}_{1}}}\sum_{s=0}^{M-1}\sum_{p=0}^{s}\frac{\left(-1\right)^{s+p}\binom{M-1}{s}\binom{s}{p}} {\frac{(p+1)}{\tilde{\Omega}_{1}}+\frac{{\gamma}_{0}}{\tilde{\Omega}_{0}}}.
\end{align}
Secondly, using (\ref{CDF_y1}) and (\ref{J1_closed}), the closed-form expression of outage probability of $K_{2}$ can be calculated as
\begin{align}
    K_{2}=&\left(1-K_{1}\right)\left[\int_{0}^{\gamma_{s}} f_{Y_{1}}\left(y\right) \mathrm{d} y \right]\\
    \nonumber
    =&\left(1-K_{1}\right)\left[M \sum_{s=0}^{M-1} \frac{\left(-1\right)^{s}}{1+s}\binom{M-1}{s} \left(1-e^{-\frac{\gamma_{s}}{\tilde{\Omega}_{1}}}\right)^{s+1}\right].
\end{align}
Finally, the closed form expression of $\mathrm{P}_{A}^{QoS}$ is obtained by writing $K_{1}$ and $K_{2}$ in (\ref{j1_and_j2}).

\bibliographystyle{IEEEtran}
\bibliography{noma_uplink_sic}

\begin{thebibliography}{10}
\providecommand{\url}[1]{#1}
\csname url@samestyle\endcsname
\providecommand{\newblock}{\relax}
\providecommand{\bibinfo}[2]{#2}
\providecommand{\BIBentrySTDinterwordspacing}{\spaceskip=0pt\relax}
\providecommand{\BIBentryALTinterwordstretchfactor}{4}
\providecommand{\BIBentryALTinterwordspacing}{\spaceskip=\fontdimen2\font plus
\BIBentryALTinterwordstretchfactor\fontdimen3\font minus
  \fontdimen4\font\relax}
\providecommand{\BIBforeignlanguage}[2]{{%
\expandafter\ifx\csname l@#1\endcsname\relax
\typeout{** WARNING: IEEEtran.bst: No hyphenation pattern has been}%
\typeout{** loaded for the language `#1'. Using the pattern for}%
\typeout{** the default language instead.}%
\else
\language=\csname l@#1\endcsname
\fi
#2}}
\providecommand{\BIBdecl}{\relax}
\BIBdecl

\bibitem{9802823}
T.~M. Hoang, L.~T. Dung, B.~C. Nguyen, X.~H. Le, X.~N. Tran, and T.~Kim,
  ``Outage {Probability} and {Throughput} of {Mobile} {Multiantenna}
  {UAV}-assisted {FD-NOMA} {Relay} {System} {With} {Imperfect} {CSI},''
  \emph{IEEE Systems Journal}, pp. 1--12, 2022.

\bibitem{elhalawany2021spectrum}
B.~M. ElHalawany, A.~A.~A. El-Banna, Q.-V. Pham, K.~Wu, and E.~M. Mohamed,
  ``Spectrum sharing in cognitive-radio-inspired {NOMA} systems under imperfect
  {SIC} and cochannel interference,'' \emph{IEEE Systems Journal}, vol.~16,
  no.~1, pp. 1540--1547, 2021.

\bibitem{gao2018cognitive}
Y.~Gao, H.~He, Z.~Deng, and X.~Zhang, ``Cognitive radio network with
  energy-harvesting based on primary and secondary user signals,'' \emph{IEEE
  Access}, vol.~6, pp. 9081--9090, 2018.

\bibitem{hu2018full}
F.~Hu, B.~Chen, and K.~Zhu, ``Full spectrum sharing in cognitive radio networks
  toward {5G}: A survey,'' \emph{IEEE Access}, vol.~6, pp. 15\,754--15\,776,
  2018.

\bibitem{7273963}
Z.~Ding, P.~Fan, and H.~V. Poor, ``Impact of user pairing on {5G} nonorthogonal
  multiple-access downlink transmissions,'' \emph{IEEE Transactions on
  Vehicular Technology}, vol.~65, no.~8, pp. 6010--6023, 2016.

\bibitem{do2020performance}
D.-T. Do, M.-S. Van~Nguyen, F.~Jameel, R.~J{\"a}ntti, and I.~S. Ansari,
  ``Performance evaluation of relay-aided {CR-NOMA} for beyond {5G}
  communications,'' \emph{IEEE Access}, vol.~8, pp. 134\,838--134\,855, 2020.

\bibitem{li2022physical}
M.~Li, H.~Yuan, C.~Maple, W.~Cheng, and G.~Epiphaniou, ``Physical {Layer}
  {Security} analysis of {Cognitive} {NOMA} {Internet} of {Things}
  {Networks},'' \emph{IEEE Systems Journal}, 2022.

\bibitem{lv2016application}
L.~Lv, Q.~Ni, Z.~Ding, and J.~Chen, ``Application of non-orthogonal multiple
  access in cooperative spectrum-sharing networks over nakagami-$m$ fading
  channels,'' \emph{IEEE Transactions on Vehicular Technology}, vol.~66, no.~6,
  pp. 5506--5511, 2016.

\bibitem{arzykulov2018outage}
S.~Arzykulov, T.~A. Tsiftsis, G.~Nauryzbayev, and M.~Abdallah, ``Outage
  performance of cooperative underlay {CR-NOMA} with imperfect {CSI},''
  \emph{IEEE Communications Letters}, vol.~23, no.~1, pp. 176--179, 2018.

\bibitem{lv2018noma}
L.~Lv, L.~Yang, H.~Jiang, T.~H. Luan, and J.~Chen, ``When {NOMA} meets
  multiuser cognitive radio: Opportunistic cooperation and user scheduling,''
  \emph{IEEE Transactions on Vehicular Technology}, vol.~67, no.~7, pp.
  6679--6684, 2018.

\bibitem{lv2017design}
L.~Lv, J.~Chen, Q.~Ni, and Z.~Ding, ``Design of cooperative non-orthogonal
  multicast cognitive multiple access for {5G} systems: User scheduling and
  performance analysis,'' \emph{IEEE Transactions on Communications}, vol.~65,
  no.~6, pp. 2641--2656, 2017.

\bibitem{liu2022new}
H.~Liu, Z.~Bai, H.~Lei, G.~Pan, K.~J. Kim, and T.~A. Tsiftsis, ``A new rate
  splitting strategy for uplink {CR-NOMA} systems,'' \emph{IEEE Transactions on
  Vehicular Technology}, vol.~71, no.~7, pp. 7947--7951, 2022.

\bibitem{xiang2019physical}
Z.~Xiang, W.~Yang, G.~Pan, Y.~Cai, and Y.~Song, ``Physical layer security in
  cognitive radio inspired {NOMA} network,'' \emph{IEEE Journal of Selected
  Topics in Signal Processing}, vol.~13, no.~3, pp. 700--714, 2019.

\bibitem{hoang2022outage}
T.~M. Hoang, B.~C. Nguyen, X.~H. Le, X.~N. Tran, T.~Kim \emph{et~al.}, ``Outage
  probability and throughput of mobile multiantenna {UAV}-assisted {FD-NOMA}
  relay system with {Imperfect} {CSI},'' \emph{IEEE Systems Journal}, 2022.

\bibitem{caceres2022theoretical}
F.~M. Caceres, S.~Kandeepan, and S.~Sun, ``Theoretical analysis of hybrid {SIC}
  success probability under rayleigh channel for uplink {CR-NOMA},'' \emph{IEEE
  Transactions on Vehicular Technology}, 2022.

\bibitem{ding2021new}
Z.~Ding, R.~Schober, and H.~V. Poor, ``A new qos-guarantee strategy for {NOMA}
  assisted semi-grant-free transmission,'' \emph{IEEE Transactions on
  Communications}, 2021.

\bibitem{9755045}
H.~Liu, Z.~Bai, H.~Lei, G.~Pan, K.~J. Kim, and T.~Tsiftsis, ``A new rate
  splitting strategy for uplink {CR-NOMA} systems,'' \emph{IEEE Transactions on
  Vehicular Technology}, pp. 1--1, 2022.

\bibitem{ding2020unveiling}
Z.~Ding, R.~Schober, and H.~V. Poor, ``Unveiling the importance of {SIC} in
  {NOMA} systems—part 1: State of the art and recent findings,'' \emph{IEEE
  Communications Letters}, vol.~24, no.~11, pp. 2373--2377, 2020.

\bibitem{ding2020unveiling1}
{Ding, Zhiguo and Schober, Robert and Poor, H Vincent}, ``Unveiling the
  importance of {SIC} in {NOMA} systems—part ii: New results and future
  directions,'' \emph{IEEE Communications Letters}, vol.~24, no.~11, pp.
  2378--2382, 2020.

\bibitem{ding2015impact}
Z.~Ding, P.~Fan, and H.~V. Poor, ``Impact of user pairing on {5G} nonorthogonal
  multiple-access downlink transmissions,'' \emph{IEEE Transactions on
  Vehicular Technology}, vol.~65, no.~8, pp. 6010--6023, 2015.

\bibitem{ding2014performance}
Z.~Ding, Z.~Yang, P.~Fan, and H.~V. Poor, ``On the performance of
  non-orthogonal multiple access in {5G} systems with randomly deployed
  users,'' \emph{IEEE Signal Processing Letters}, vol.~21, no.~12, pp.
  1501--1505, 2014.

\bibitem{zhou2018state}
F.~Zhou, Y.~Wu, Y.-C. Liang, Z.~Li, Y.~Wang, and K.-K. Wong, ``State of the
  art, taxonomy, and open issues on cognitive radio networks with {NOMA},''
  \emph{IEEE Wireless Communications}, vol.~25, no.~2, pp. 100--108, 2018.

\bibitem{7542118}
Z.~Yang, Z.~Ding, P.~Fan, and N.~Al-Dhahir, ``A general power allocation scheme
  to guarantee quality of service in downlink and uplink {NOMA} systems,''
  \emph{IEEE Transactions on Wireless Communications}, vol.~15, no.~11, pp.
  7244--7257, 2016.

\bibitem{1413621}
Y.~Chen and C.~Tellambura, ``Performance analysis of maximum ratio transmission
  with imperfect channel estimation,'' \emph{IEEE Communications Letters},
  vol.~9, no.~4, pp. 322--324, 2005.

\bibitem{gu2003performance}
D.~Gu and C.~Leung, ``Performance analysis of transmit diversity scheme with
  imperfect channel estimation,'' \emph{Electronics letters}, vol.~39, no.~4,
  pp. 402--403, 2003.

\bibitem{7510916}
F.~Liu, P.~Mähönen, and M.~Petrova, ``Proportional fairness-based power
  allocation and user set selection for downlink {NOMA} systems,'' in
  \emph{2016 IEEE International Conference on Communications (ICC)}, 2016, pp.
  1--6.

\bibitem{men2016performance}
J.~Men, J.~Ge, and C.~Zhang, ``Performance analysis of nonorthogonal multiple
  access for relaying networks over {Nakagami-$m$} fading channels,''
  \emph{IEEE Transactions on Vehicular Technology}, vol.~66, no.~2, pp.
  1200--1208, 2016.

\end{thebibliography}
\end{document}